\documentclass[twocolumn,preprintnumbers,amsmath,amssymb]{revtex4}

\usepackage{graphicx}
\usepackage{dcolumn}
\usepackage{bm}
\newcommand{\bhat}[1]{\hat{\boldsymbol{#1}}}

\begin{document}


\title{Implementation of a symmetric surface electrode ion trap with field compensation using a modulated Raman effect}

\author{D. T. C. Allcock}
 \email{d.allcock@physics.ox.ac.uk}
\author{J. A. Sherman}
\author{D. N. Stacey}
\author{A. H. Burrell}
\author{M. J. Curtis}
\author{G. Imreh}
\author{N. M. Linke}
\author{D. J. Szwer}
\author{S. C. Webster}
\author{A. M. Steane}
\author{D. M. Lucas}
\affiliation{Department of Physics, University of Oxford, Clarendon Laboratory, Parks Road, Oxford OX1 3PU, United Kingdom}

\date{\today}

\begin{abstract}
We describe the fabrication and characterization of a new surface-electrode Paul ion trap designed for experiments in scalable quantum information processing with Ca$^+$.  A notable feature is a symmetric electrode pattern which allows rotation of the normal modes of ion motion, yielding efficient Doppler cooling with a single beam parallel to the planar surface.  We propose and implement a technique for micromotion compensation in all directions using an infrared repumper laser beam directed into the trap plane.  Finally, we employ an alternate repumping scheme that increases ion fluorescence and simplifies heating rate measurements obtained by time-resolved ion fluorescence during Doppler cooling.
\end{abstract}

\maketitle
The ion trap is a leading candidate technology for realizing a quantum information processor;  the required steps of initialization, processing, qubit transport and readout have been demonstrated in a single device \cite{Hom09}.  These operations are currently at or approaching fault-tolerant 
thresholds \cite{Mye08,Ben08}. A current practical challenge is the development of ion traps capable of storing and precisely manipulating a substantial number of ions \cite{Kie02}.  Surface-electrode, or planar, designs \cite{Chi05} with arbitrary electrode arrangements are easily fabricated and more easily integrated with on-chip control electronics and optics.  However the best choices for materials and fabrication methods remain unclear.

We demonstrate the fabrication and evaluation of a gold-on-silica planar ion trap with a 5~$\mu$m feature size and ion-to-electrode separation of 150~$\mu$m.  We also describe new Doppler-cooling and trap optimization techniques specific to planar traps and generalizable to many commonly studied ion species.

\section{\label{sec:volts}Trap Design}

\subsection{Geometry}
\raggedbottom

\begin{figure}
\includegraphics[width=.9\columnwidth]{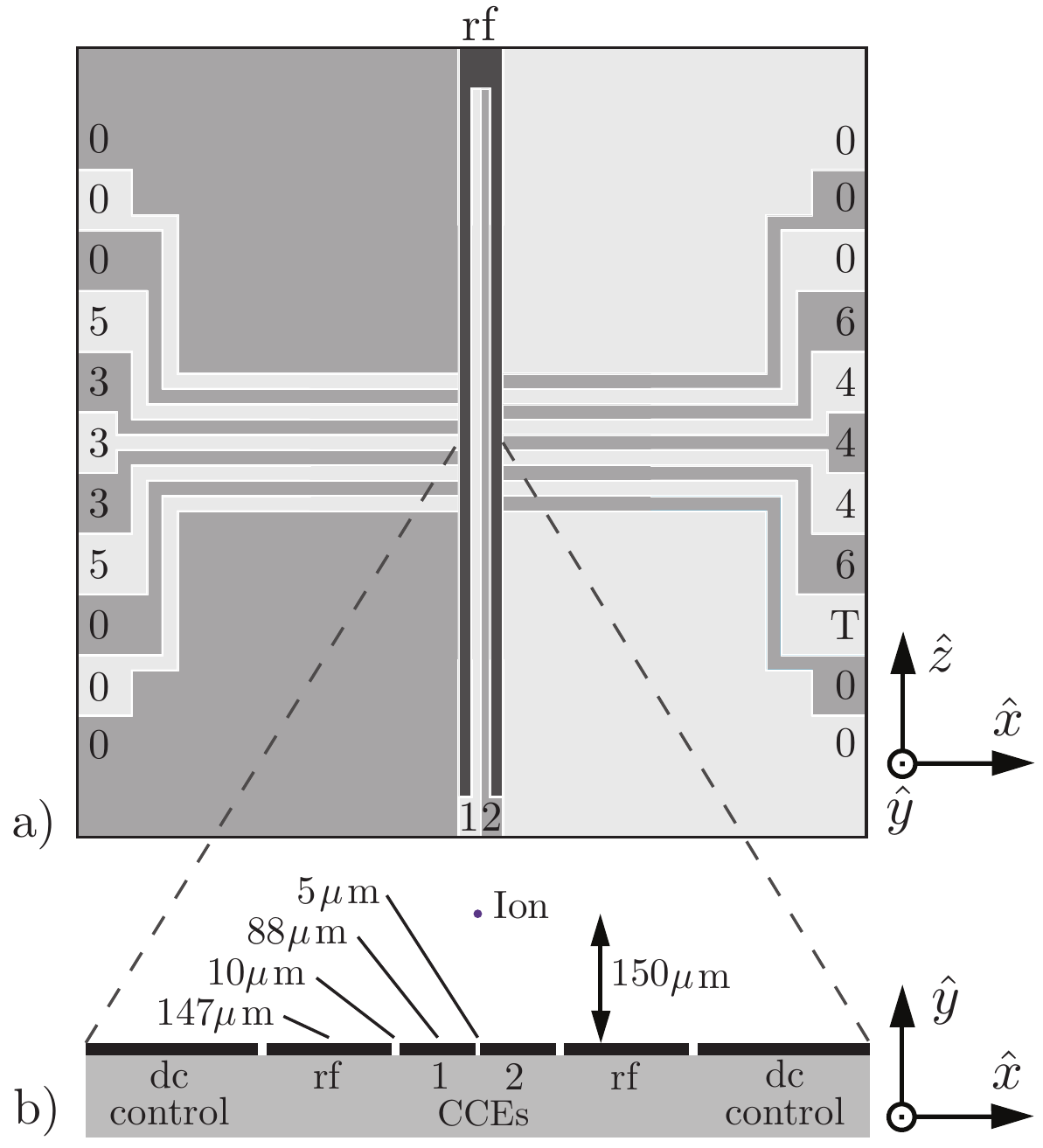}
\caption{\label{fig:design} Trap design. (a) Plan view. The rf rails are dark gray and the dc control electrodes are alternately picked out in light/medium gray for clarity.  The labels refer to the different voltages, V$_1\ldots$V$_6$, we apply. Other dc electrodes are grounded (0) except `T' which is used for secular frequency measurements as described in Section \ref{sec:trap}. (b) End view showing electrode and gap widths, and ion position.}
\end{figure}

\begin{figure}
\includegraphics[width=.9\columnwidth]{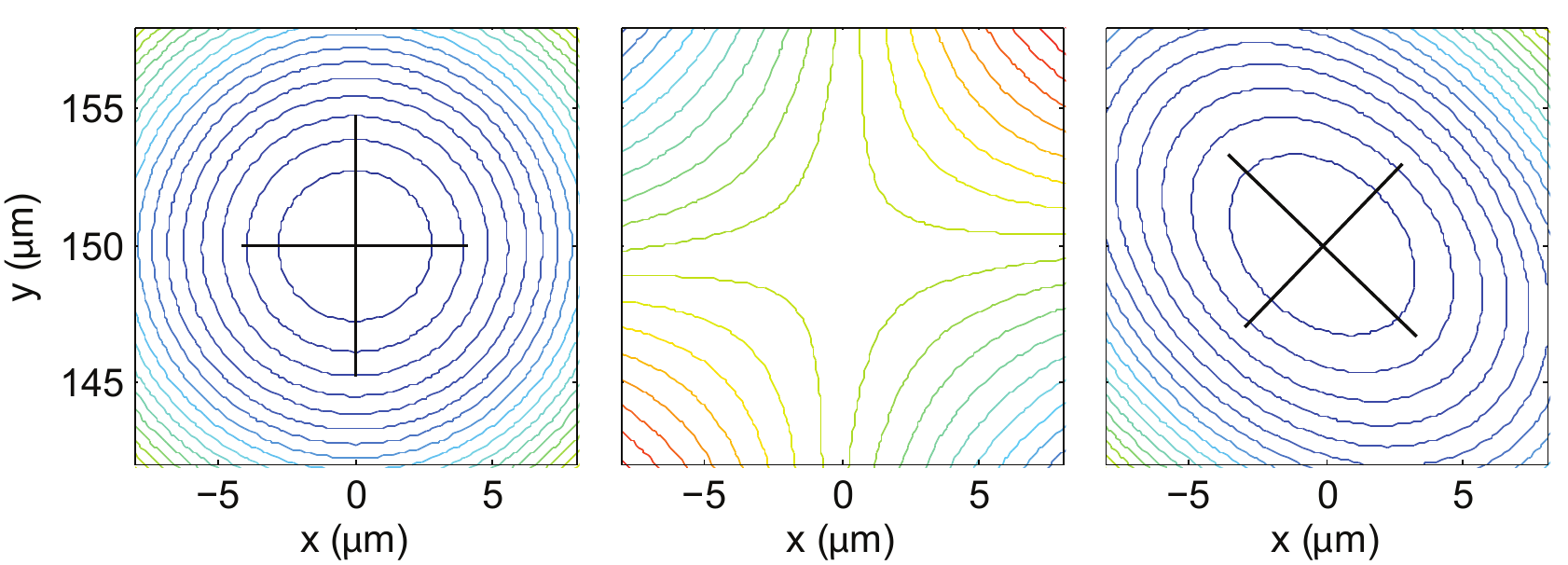}
\caption{\label{fig:pots} The rf pseudopotential (left), the rotation quadrupole potential (center) and the superposition of the two (right).  The straight lines show the axes of the normal modes of the secular ion motion.}
\end{figure}

We selected a linear trap geometry with electrodes symmetrical about the trap's axial ($\bhat{z}$) direction.  This geometry is uncommon as the symmetry causes the radial principal axes to lie in the $\bhat{x}$ and $\bhat{y}$ directions (see Fig.~\ref{fig:design}a).  If the Doppler cooling lasers are to pass across the trap's surface without striking it then they will have no projection in the $\bhat{y}$ direction, leaving that mode uncooled. Asymmetric designs \cite{Lab08a, Sei06} rotate these modes so they both couple to a beam parallel to the surface.  We choose instead to split the central axial control electrode (between the radio frequency (rf) electrodes) to create a `six-wire' geometry (see Fig.~\ref{fig:design}b) \cite{Ste06}. We use these two electrodes and the outer segmented electrodes to place a static quadrupole (oriented at 45$^{\circ}$ w.r.t.\ the cardinal axes) over the cylindrically symmetric rf pseudopotential to break the degeneracy of the normal modes and ensure rotation of their axes to achieve efficient Doppler cooling (see Fig. \ref{fig:pots}).  One advantage of this approach is that it does not require one of the rf electrodes to be significantly larger than the other, allowing the use of smaller electrodes, leading to less capacitive coupling and lower losses in the rf trap drive --- especially important when using a high-loss semiconductor substrate \cite{Luc09}. Symmetrical rf electrodes also simplify selection of dc control voltages in complicated arrangements, such as junctions \cite{Ami08}.  Finally the gap between the two center electrodes directly below the ion could be used to provide optical access to the ion.  This symmetric trap does have the disadvantage that it is not as tight or deep as the optimal `four-wire' geometry \cite{Wes08}.

The trap has 11 pairs of outer dc control electrodes. We presently only control the central five pairs, grounding the others.  The ions are 274~$\mu$m from the control electrodes, which are 145~$\mu$m wide in the $\bhat{z}$ direction.  The ratio of electrode width to ion distance is 0.53 which is of the appropriate order of magnitude for optimal transport control \cite{Rei06}.

Gaps between electrodes in the trapping region are 10~$\mu$m as these can be reliably fabricated and should give a breakdown voltage of at least 300~V \cite{LabTh}.  Directly under the ion the gap is reduced to 5~$\mu$m to reduce the exposed dielectric as there is no high rf voltage across this gap. Far ($>1500~\mu$m) from the ion the gaps grow to 15--25~$\mu$m to improve fabrication yield and reduce capacitive load on the rf drive. Another design feature is the extension of the rf rails across most of the chip to reduce end effects which cause axial rf field gradients, driving micromotion.

\subsection{Operating Parameters}

We model our trap using numeric boundary element software (Charged Particle Optics) and verify the model using an analytical method \cite{Wes08}.  Using the model we first calculate the position and strength of the radially confining rf pseudopotential which gives us the trapped position of the ion.  We then calculate the dc potential around this point due to a unit voltage applied to the $i$th dc control electrode.  We fit a quadratic function to the calculated dc potential, 
\begin{equation}
\label{eqn:pot}
\Phi_i=\alpha_{xi}x^2+\alpha_{yi}y^2+\alpha_{zi}z^2+\beta_{xi}x+\beta_{yi}y+\beta_{zi}z+C_i.
\end{equation}
This gives us a set of quadratic {\boldmath$\alpha$} and linear {\boldmath$\beta$} coefficients for each electrode $i$ (the constants $C_i$ are ignored as they do not affect the analysis). We now wish to {\em specify} a particular potential, for example,
\begin{equation}
\Phi=\alpha_{x}x^2+\alpha_{y}y^2+\alpha_{z}z^2+\beta_{x}x+\beta_{y}y+\beta_{z}z
\end{equation}
and find what electrode voltages $V_1\ldots V_6$ are necessary to obtain this potential. The equation
\begin{equation}
\label{eqn:volt}
\left(\begin{array}{c}
	\beta_{x}\\
	\beta_{y}\\
	\beta_{z}\\
  \alpha_{x}\\
	\alpha_{y}\\
	\alpha_{z}
	\end{array}\right)
	=
\left(\begin{array}{llll}
	\beta_{x1} & \beta_{x2} & \cdots & \beta_{x6}\\
	\beta_{y1} & \beta_{y2} & \cdots & \beta_{y6}\\
	\beta_{z1} & \beta_{z2} & \cdots & \beta_{z6}\\
  \alpha_{x1} & \alpha_{x2} & \cdots & \alpha_{x6}\\
	\alpha_{y1} & \alpha_{y2} & \cdots & \alpha_{y6}\\
	\alpha_{z1} & \alpha_{z2} & \cdots & \alpha_{z6}
\end{array}\right)
\times
\left(\begin{array}{c}
	V_1\\
	V_2\\
	V_3\\
	V_4\\
	V_5\\
	V_6
\end{array}\right)
\end{equation}
can be used to find the required voltages. The number of independent voltages is typically fewer than six, since symmetry or Laplace's equation (which ensures $\alpha_{x}+\alpha_{y}+\alpha_z=0$) reduces the number of degrees of freedom.

We define four sets of operating parameters, each with a required form of the potential given by the {\boldmath$\alpha$} and {\boldmath$\beta$} coefficients in Table \ref{tab:efields}.  The four operating parameter sets are: the `endcap' set confining the ion axially, the `tilt' set controlling the orientation of the radial normal modes, and sets for $\bhat{x}$ and $\bhat{y}$ micromotion compensation.  The latter two apply linear fields at the site of the ion in $\bhat{x}$ and $\bhat{y}$ respectively to trim out stray fields that move the ion off the pseudopotential null and increase the driven micromotion \cite{Ber98}.  To simplify operation of the trap we choose the constraints such that adjusting one parameter minimally affects the others.  This is especially important in the case of the endcap set as the radial curvature it induces must be radially symmetric or it will reduce the radial normal mode tilt angle.  Applying the `tilt' potential weakly couples the $\bhat{x}$ and $\bhat{y}$ micromotion compensation.  The chosen tilt voltages break the radial mode degeneracy without overly weakening either of them or significantly reducing the trap depth. We ignore quadratic cross terms (eg. $xy$) in Eq.~\ref{eqn:pot}: these are zero by symmetry for the `endcap' and $\bhat{y}$-compensation bases, and negligibly small for the $\bhat{x}$-compensation basis.  For the tilt potential we use an alternate basis [$\bhat{x}'=(\bhat{x}+\bhat{y})/\sqrt{2}$, $\bhat{y}'=(\bhat{x}-\bhat{y})/\sqrt{2}$ and $\bhat{z}'=\bhat{z}$] so that again the quadratic cross-terms vanish by symmetry.  In this model the tilt angle would always be 45$^{\circ}$ as soon as the radial mode degeneracy was lifted; in practice, end effects modify this angle.  The angle can be estimated (see Table. \ref{tab:freqa}) from a fit to the full potential (sum of rf pseudopotential, `endcap' potential and `tilt' potential).

 The endcap voltages shown in Table~\ref{tab:vbasis} give an axial secular frequency of 500~kHz.  The radial secular frequency is independently set by the rf voltage applied to the rf electrodes.  At our typical rf voltage, the `tilt' voltages in Table~\ref{tab:vbasis} give radial secular frequencies of 2.29, 3.33~MHz and an estimated tilt angle of 42$^{\circ}$.

\begin{table}5\caption{\label{tab:efields}The coefficients of the potentials for our four voltage bases. $\beta$ is in units of Vm$^{-1}$ and $\alpha$ in units of Vm$^{-2}$. For the `tilt' parameters we use the alternate basis $\bhat{x}'=(\bhat{x}+\bhat{y})/\sqrt{2}$, $\bhat{y}'=(\bhat{x}-\bhat{y})/\sqrt{2}$ and $\bhat{z}'=\bhat{z}$.}
\begin{ruledtabular}
\begin{tabular}{ccccccc}
 & $\beta_{x}$ & $\beta_{y}$ & $\beta_{z}$ & $\alpha_{x}$ & $\alpha_{y}$ & $\alpha_{z}$\\
\hline
Endcap & 0 & 0 & 0 & -1.02$\times10^{6}$ & -1.02$\times10^{6}$ & 2.05$\times10^{6}$\\
$\bhat{x}$-Comp & 1 & 0 & 0 & 0 & 0 & 0\\
$\bhat{y}$-Comp & 0 & 1 & 0 & 0 & 0 & 0\\
\hline
 & $\beta_{x'}$ & $\beta_{y'}$ & $\beta_{z'}$ & $\alpha_{x'}$ & $\alpha_{y'}$ & $\alpha_{z'}$\\
\hline
Tilt & 0 & 0 & 0 & 1.00$\times10^{7}$ & -1.00$\times10^{7}$ & 0\\
\end{tabular}
\end{ruledtabular}
\end{table}

\begin{table}
\caption{\label{tab:vbasis}The calculated voltage bases (in mV).}
\begin{ruledtabular}
\begin{tabular}{ccccccc}
 & V$_{1}$ & V$_{2}$ & V$_{3}$ & V$_{4}$ & V$_{5}$ & V$_{6}$\\
\hline
Endcap & -1040 & -1040 & -3152 & -3152 & -235 & -235\\
$\bhat{x}$-Comp & 0 & 0 & -0.95 & 0.95 & -0.95 & 0.95\\
$\bhat{y}$-Comp & 0.92 & 0.92 & 1.86 & 1.86 & 5.02 & 5.02\\
Tilt & -886 & 929 & 1117 & -1030 & 1117 & -1030\\
\end{tabular}
\end{ruledtabular}
\end{table}

\section{Experimental Apparatus}

\subsection{Trap fabrication}

\begin{figure}
\includegraphics[width=1\columnwidth]{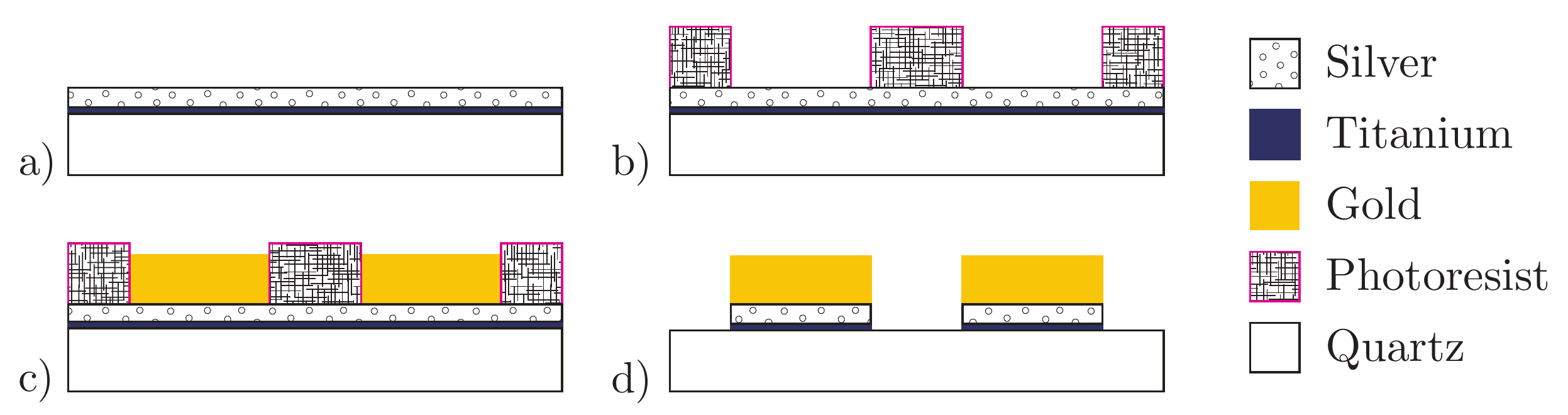}
\caption{\label{fig:trapfab} Fabrication stages (see text).}
\end{figure}

\begin{figure}
\includegraphics[width=.6\columnwidth]{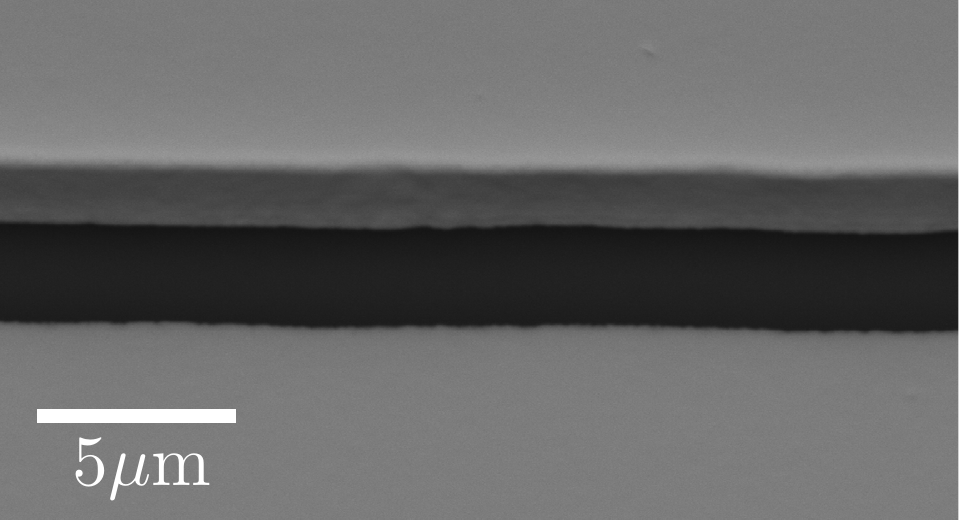}
\caption{\label{fig:semimg} SEM (JEOL - JSM-6480) image of an inter-electrode gap taken at 55$^\circ$ to the plane to show the gold thickness.}
\end{figure}

Our fabrication process is based on that of I. Chuang's group at MIT \cite{LabTh} but uses the electroplating process developed in \cite{Kou04}.  We fabricated the trap on a \mbox{$10\times10\times0.5$~mm} polished single-crystal quartz substrate.  The substrate is thoroughly cleaned in piranha etch (10 parts H$_{2}$SO$_{4}$, 1 part H$_{2}$O$_{2}$) at 100$^{\circ}$C for 3 minutes then rinsed in high quality ($<$0.0001\% residue) de-ionized (DI) water.  The first stage of fabrication is depositing the silver seed layer.  The substrate is glow-discharge cleaned and a 15~nm titanium layer is evaporatively deposited to ensure good adhesion before evaporating on 200~nm of silver (Fig.~\ref{fig:trapfab}a).  The substrate is then cleaned in acetone, isopropyl alcohol (IPA) and DI water followed by a bake at 115$^{\circ}$C to dry.  To promote good photoresist adhesion a primer (20\% hexamethyldisilazane (HMDS), 80\% propylene glycol monomethyl ether acetate) is puddled on the substrate for 10s before being spun off at 5000~rpm for 45~seconds. The photoresist (SPR 220-7.0) is puddled onto the substrate and spun off at 5000~rpm for 45 seconds to give a nominal thickness of 5~$\mu$m. After being spun, the photoresist is baked for 2 minutes at 115$^{\circ}$C before resting at least 30 minutes to rehydrate the photoresist.  A mask aligner is used with chrome on glass photomask (JD Phototools) for the photolithography of the trap structure.  This is done in two stages.  The first exposes and removes the edge-bead caused by photoresist surface tension to ensure good mask contact and to prevent undesirable regions of bare quartz on the trap.  The second defines the trap structure (Fig.~\ref{fig:trapfab}b).  After each stage there is a 2 minute development (using Microposit MF CD-26) and DI water rinse.  A fully bright sulphite gold electroplating process (Metalor ECF 63) plates the electrodes up within the photoresist mask (Fig.~\ref{fig:trapfab}c).  Plating is carried out using 200~ml of solution in a beaker on a stirring hotplate.  The trap and anode are suspended in the beaker facing each other at a distance of a few cm.  The anode is pure platinum foil with an area of $\sim$500~mm$^{2}$.  Plating was carried out at 55$^{\circ}$C in a constantly stirred solution with 3~mA of current which gave us $\sim$2.7~$\mu$m of gold in 30 minutes.  After electroplating the photoresist is removed using acetone, IPA and DI water. Then the silver seed layer is etched using NH$_{4}$OH:H$_{2}$O$_{2}$:H$_{2}$O (1:1:4) for 15 seconds.  At this point the trap is inspected and the gold thickness measured using a scanning electron microscope (SEM, see Fig.~\ref{fig:semimg}).  The surface quality obtained is mirror-like and no surface features could be found by the SEM which indicates an upper bound on surface roughness of around 50~nm, in agreement with atomic force microscope studies of similar devices\cite{Kou04}.  If the trap passes inspection the titanium is etched using HF:H$_{2}$O (1:10) for 15 seconds (Fig.~\ref{fig:trapfab}d). Final cleaning is carried out using acetone, IPA and then DI water.  Any remaining organic residue is then removed by placing the trap under an O$_{3}$-producing mercury lamp for several minutes.
\par

\begin{figure}
\includegraphics[width=.9\columnwidth]{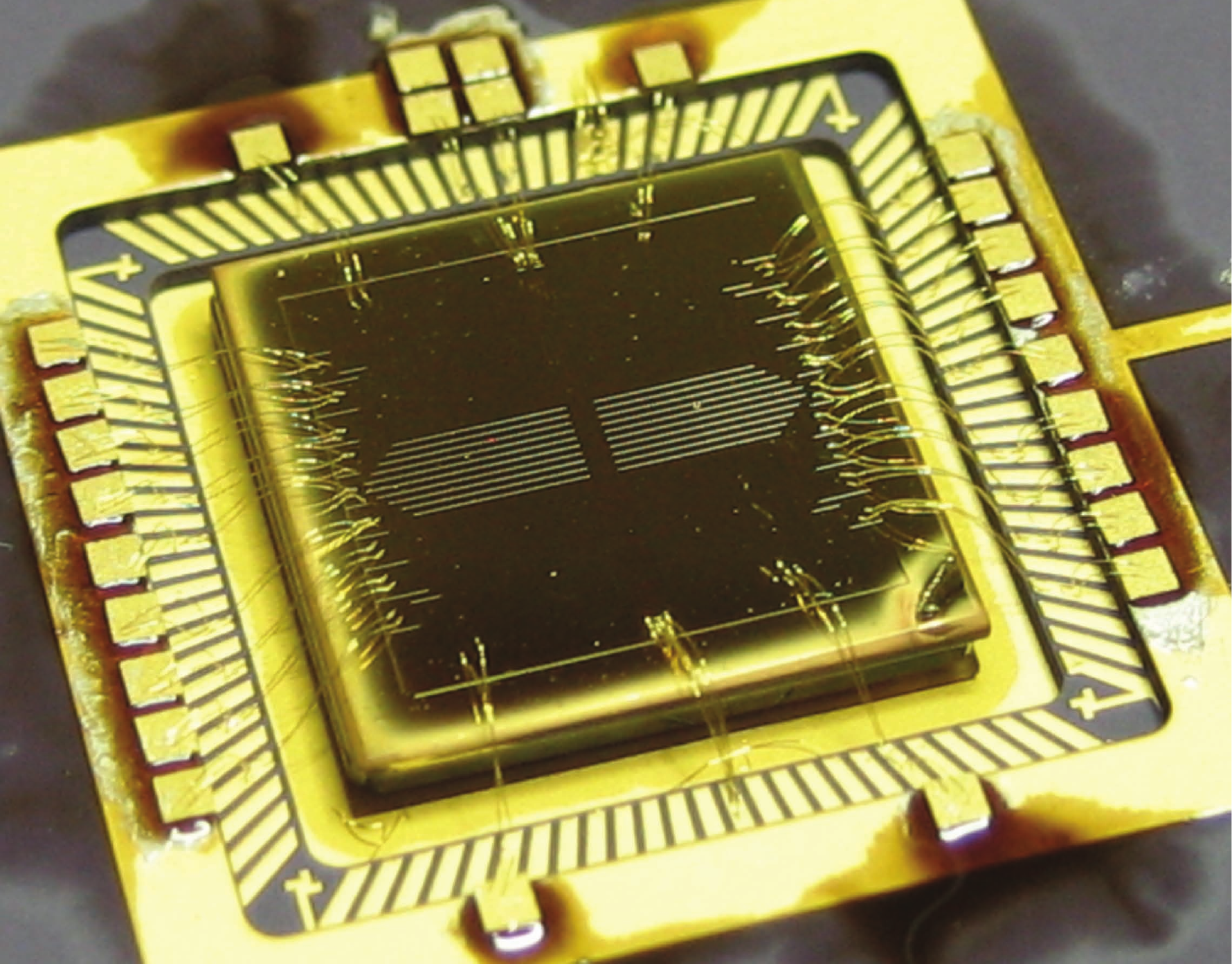}
\caption{\label{fig:packtrap} Final packaged trap.  The gold squares are the single layer capacitors.}
\end{figure}

The completed trap (see Fig.~\ref{fig:packtrap}) was glued to a 100-pin Kyocera ceramic pin grid array (CPGA) package with Epo-Tek 353ND epoxy via a 1mm fused silica spacer.   Each electrode is wire bonded to the package with two bonds except the rf which has five bonds.  820~pF single layer capacitors (American Technical Ceramics) are mounted directly to the package to provide a low-impedance path to ground for any rf coupled onto the dc electrodes.  These are attached to the outer gold ring of the CPGA using Epo-Tek H20E conducting epoxy.  353ND epoxy is then applied around each capacitor to prevent them being shorted by calcium deposition from the oven used to load the trap.  The outer ring is grounded and the capacitors wire-bonded to each electrode's pad on the package with two bonds.  Two capacitors are used on each of the center control electrodes as they have higher capacitance to the rf rails.

\subsection{Vacuum System}
The socket for the trap package is mounted in a 114~mm diameter CF flange octagon (Kimball Physics).  This socket is comprised of two UHV-compatible plastic (Vespel SP-3) plates with pin receptacles confined between them as described in \cite{StiTh}.  The receptacles for dc voltages are crimped to a 25-way kapton-coated ribbon cable connected to a D-sub vacuum feedthrough.  The receptacle for the rf voltage goes via a single kapton coated cable to a 1-pin feedthrough on the octagon.  Two 20~l/s ion pumps and a non-evaporable getter pump give a pressure, after a 190$^{\circ}$C bakeout, of $\sim3\times10^{-10}$ Torr which we believe is limited by a faulty UHV valve.  The calcium oven is a stainless steel tube containing calcium metal with a $\sim$0.3~mm diameter hole in the side.  This is spot-welded to a two-pin feedthrough and resistively heated to produce a neutral Ca beam.  The oven feedthrough is mounted on the octagon such that the neutral beam is perpendicular to the lasers.  There is a glass viewport opposite the oven to allow it to be aligned optically so that the neutral beam is almost parallel to the trap surface to reduce calcium deposition on the electrodes and possible shorting.  Above the trap, at a distance of $\sim$30~mm, is a fused-silica imaging window.  It has an anti-reflection (AR) coating on the outside and a conductive indium tin oxide (ITO) coating on the inside (4.7~M$\Omega$ from center of glass to flange) to prevent charging.  Transmission is 87\% at 397~nm.  The side windows are fused silica in 34~mm CF flanges with a broad-band AR coating on both sides.  Ion fluorescence is collected by a lens with an NA of 0.29, filtered to remove scatter from infrared lasers and then either counted using a PMT (overall efficiency of 0.24\%) or imaged with a CCD camera (1.7~$\mu$m resolution).  Field coils are provided in three orthogonal directions to allow a B-field of up to 8~G to be applied in any direction.

\subsection{Voltage Supplies}
The rf trap drive is produced by a Stanford Research Systems DS345 Synthesizer, amplified by a Mini-Circuits ZHL-1-2W, passed through a directional coupler (Minicircuits ZDC-20-3) for monitoring VSWR and finally stepped up with an inductively coupled helical resonator.  The loaded resonant frequency of the resonator is 25.8 MHz.  A two-turn pickup coil around the rf vacuum feedthrough allows us to monitor the trap voltage. Dc voltages are calculated using a Labview program which takes desired tilt, axial frequency and micromotion compensation parameters and produces the required voltages with a 16-bit DAC (Measurement Computing PCI DAC67703).  The output from the DAC is filtered using a C-R-C $\Pi$ filter (R=1~M$\Omega$, C=0.1~$\mu$F) just before the dc vacuum feedthrough.

\subsection{Lasers}

\begin{figure}
\includegraphics[width=1\columnwidth]{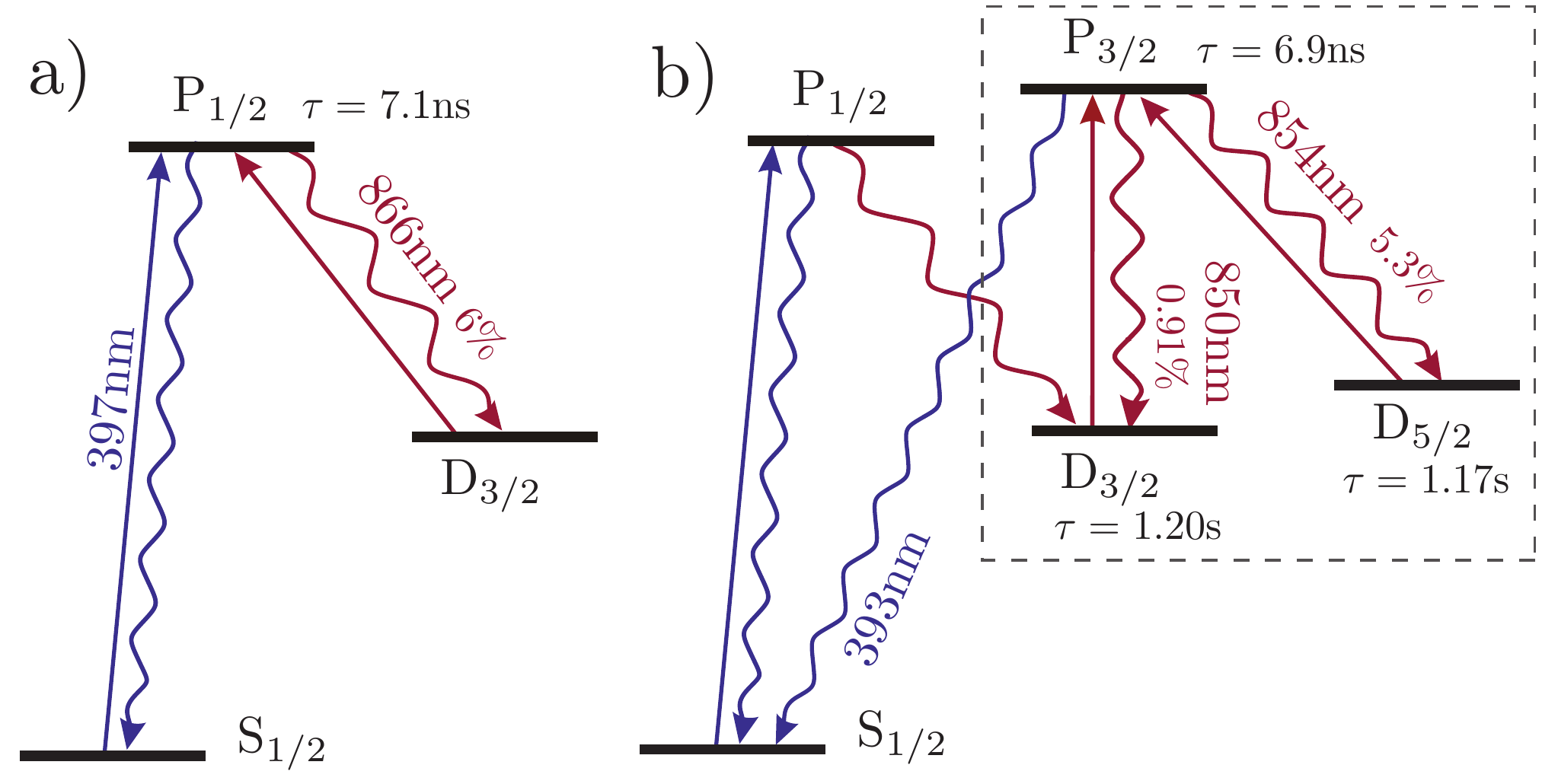}
\caption{\label{fig:calevs} Doppler cooling with (a) 866~nm repumping and (b) 850~nm/854~nm repumping.  The percentages are the branching ratios of the three repump transitions.  In (b), because there is no laser linking the boxed level system with the S$_{1/2}$--P$_{3/2}$ system, and the P$_{1/2}\rightarrow$D$_{3/2}$ branching ratio is small, the S$_{1/2}$--P$_{3/2}$ system behaves like a quasi two-level atom.}
\end{figure}

The laser wavelengths required for these experiments (see Fig.~\ref{fig:calevs}) are 397~nm for Doppler cooling and 850~nm, 854~nm, and 866~nm for D-state repumping.  In addition 423~nm and 389~nm lasers are used for two-stage photoionization \cite{Ox04}.  All lasers are Toptica DL100 extended cavity diode lasers (except the non-resonant 389~nm which has no extended cavity) and delivered to the experimental table through three polarization-maintaining single-mode fibers, one for 397~nm, one for the infrared lasers and one for the photoionization lasers.  Resonant Ca$^+$ lasers are locked to low-drift optical cavities.  All lasers are switched using double-pass AOMs except the photoionization beams which are switched with a shutter.

All the lasers are superimposed and pass parallel to the trap's surface in the $\bhat{x}+\bhat{z}$ direction.  An auxiliary 866~nm beam for micromotion compensation passes through the front window and is reflected off the trap's surface (see Fig.~\ref{fig:TrapDiag}).

\section{\label{sec:trap}Trapping}

\begin{figure}
\includegraphics[width=0.7\columnwidth]{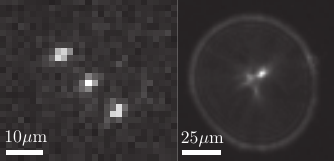}
\caption{\label{fig:3ions} Three ions in the trap (left) and a single ion with the imaging system focused on the \textit{image} of the ion in the electrodes (right).}
\end{figure}

Up to three ions (see Fig.~\ref{fig:3ions}) have been successfully trapped and crystallized.  A ratio of fluorescence signal to background scatter of over 100 is observed for a single ion. We can focus the imaging system on either the ion itself or on its reflection in the gold electrodes.  The reflection increases the fluorescence collection efficiency by up to  $\approx$50\% compared to a non-surface-electrode trap studied previously.

\subsection{Ion Lifetime}

\begin{figure}
\includegraphics[width=0.8\columnwidth]{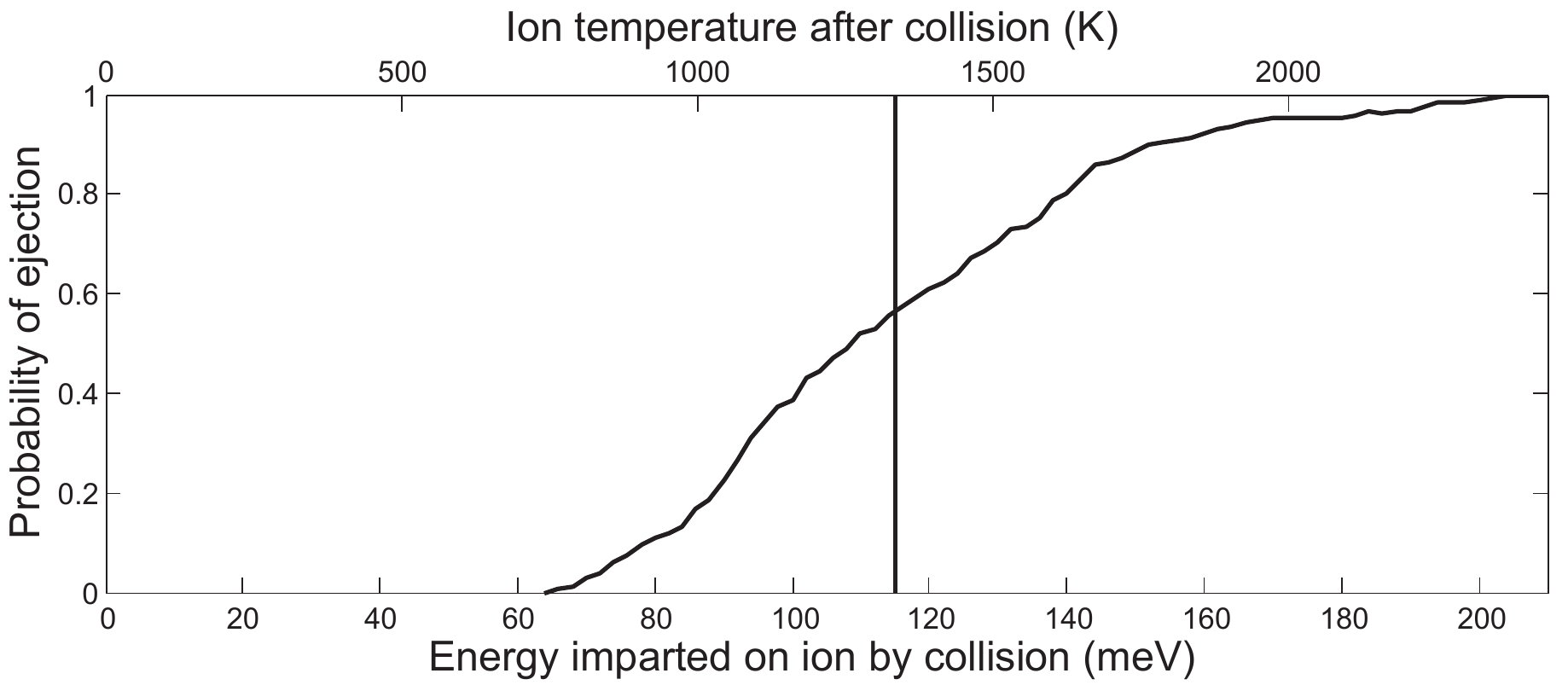}
\caption{\label{fig:losscurve} Computed probability of immediate ion loss from a collision. The vertical line is the depth of the pseudopotential.}
\end{figure}

Ion lifetime is around 5 minutes, thought to be limited by collisions resulting from the relatively high background pressure.  Similar issues with low lifetime in shallow surface traps at $\approx10^{-10}$ Torr pressures have been reported elsewhere \cite{Luc09}.  Numerical simulations of two-dimensional (in the radial plane) ion trajectories were carried out to investigate the possibility of collisional loss.  Figure~\ref{fig:losscurve} shows the calculated probability of immediate loss (within 2~$\mu$s of collision) of the ion versus the initial kinetic energy imparted to it by a collision, assuming the ion was at rest before the collision.  Each point is averaged over different rf phases and collision angles.  The results show that ion loss is possible at energies of about half the trap depth given by the calculated pseudopotential. However, the rate of `hard' collisions (where background neutrals penetrate the angular momentum barrier to collision), as calculated using Langevin theory \cite{Win98}, which impart the required collision energy is not enough to explain the observed loss rate.  One possible loss mechanism is that collisions, including more frequent `soft' collisions (which do not penetrate the angular momentum barrier), put the ion into a highly excited orbit where large Doppler shifts and poor laser beam overlap reduce cooling efficiency.  The heating rate for a hot ($\gtrsim100$~K) ion will also be greater than we measure in Section \ref{sec:heat} (where we measure ion temperatures of up to $\approx1$~K), as away from the rf null the ion is heated by noise on the rf supply and the large anharmonicities in a surface trap away from its center will cause micromotion to heat the ion \cite{Win98}.  These effects may cause the ion to be heated out of the trap more quickly than it is cooled. This is corroborated by the fact that a doubling of trap depth does not give a significant increase in trapping lifetime.  In this model the increased trap depth could be largely cancelled by the increased micromotion (see Eq.~\ref{eqn:micro}) and increased heating rate (see Fig.~\ref{fig:heatgraph}). We also see drop-outs in fluorescence of several hundred milliseconds every minute or so which could be due to collisions (they occur with the 854~nm repumper on so are not quantum jumps to the D$_{5/2}$ shelf).

\subsection{Secular Frequencies}

\begin{table}
\caption{\label{tab:freqa}Comparison of measured and modelled radial trap frequencies for different tilt factors.  $\theta_\text{model}$ is the angle of tilt for the normal mode initially perpendicular to the trap plane.  A tilt factor of 1 corresponds to $\alpha_{x'}=$1.0$\times10^{7}$~Vm$^{-2}$ (see Table \ref{tab:efields}.)}
\begin{ruledtabular}
\begin{tabular}{cccc}
Tilt factor & $f_\text{expt.}$ (MHz) & $f_\text{model}$ (MHz) & $\theta_\text{model}$\\
\hline
0 & 3.11, 3.16 & 3.12, 3.15 & 0$^{\circ}$\\
0.125 & 3.11, 3.18 & 3.10, 3.16 & 28$^{\circ}$\\
0.25 & 3.09, 3.19 & 3.08, 3.19 & 35$^{\circ}$\\
0.5 & 3.03, 3.24 & 3.03, 3.24 & 40$^{\circ}$\\
1 & 2.90, 3.34 & 2.92, 3.33 & 42$^{\circ}$\\
2 & 2.63, 3.52 & 2.70, 3.52 & 44$^{\circ}$\\
\end{tabular}
\end{ruledtabular}
\end{table}

The experimental and modelled radial secular frequencies as a function of the applied radial mode tilt are shown in Table~\ref{tab:freqa}.  These are found experimentally by applying an oscillating voltage to an electrode (marked `T' in Fig.~\ref{fig:design}) which observably heats the ion when resonant with a motional mode.  The axial secular frequency was 467 kHz compared to the model value of 500 kHz.  By working the secular frequencies back through our model of the trap we calculate an rf amplitude of 175 V, a stability parameter of $q$=0.34 and a trap depth of 115 meV.  We managed to operate the trap with rf amplitude between 223 V ($q$=0.43, $f_\text{radial}=$4.02~MHz, depth=188~meV) and 112 V ($q$=0.22, $f_\text{radial}=$1.99~MHz, depth=47~meV). The rf amplitude is the only model parameter that is fitted to the data.  The voltage amplitude before the resonator was 7.07~V implying a resonator step-up of 24.7.

\section{\label{sec:micro}Micromotion Compensation}

In a linear Paul trap any displacement of the ion off the rf null in the radial directions ($\bhat{x}$ or $\bhat{y}$) of the trap will lead to the ion experiencing driven motion, known as micromotion, due to the confining rf field \cite{Ber98}.  The expressions for the radial displacement $x_{d}$ of the ion from the rf null due to field $E_\text{dc}$ and the amplitude $x_{\mu}$ of the resulting excess micromotion are
\begin{equation}
\label{eqn:micro}
x_{\text{d}}=\frac{\text{Q}E_{\text{dc}}}{m\omega_{r}^{2}},\qquad x_{\mu}=\sqrt{2}\frac{\omega_{\text{r}}}{\Omega_{\text{rf}}}x_{\text{d}},
\end{equation}
where $\omega_{r}$ is the radial trapping frequency, $\Omega_{\text{rf}}$ is the trapping rf frequency, $m$ is the ion's mass and Q is the ion's charge.  This motion is undesirable as it will cause Doppler shifts, allows noise on the rf supply to couple to the ion and in certain circumstances can directly heat ions \cite{Win98}.  A widely used method of detecting micromotion is to measure the correlation of the arrival time of photons scattered during Doppler cooling with the rf phase \cite{Ber98}.  This correlation arises due to first order Doppler shift from the micromotion changing the apparent detuning of the cooling lasers at different points in the rf cycle. It is observed by recording a histogram of the output from a time-to-amplitude converter which measures the delay between photon arrival and a sync pulse from the trap rf supply.  The excess micromotion can then be reduced by using the compensation fields described in Section \ref{sec:volts} to shift the ion back to the rf null.

Doppler cooling beams must lie in the $(\bhat{x}-\bhat{z})$ plane of surface electrode traps to avoid striking the trap and causing scatter or charging of the substrate.  Full radial micromotion compensation cannot be carried out in this geometry using the standard correlation technique. This is because there will necessarily be a radial direction in which the ion can move $(\bhat{y})$ which is perpendicular to any laser direction and will therefore not have a Doppler shift associated with it.   

 Other methods of compensating in this direction exist, such as measuring the ion position as a function of rf voltage by translating the Doppler cooling laser \cite{Bro07} or minimizing the heating rate of the ions \cite{LabTh}.  Rf correlation is more convenient than these methods and more accurate than the first.
 
\begin{figure}
\includegraphics[width=0.9\columnwidth]{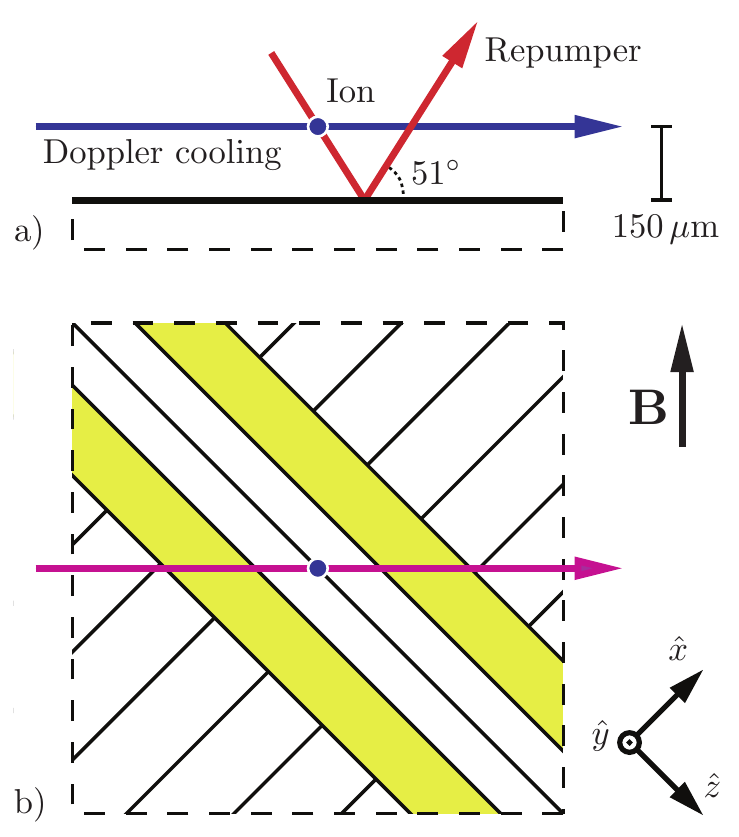}
\caption{\label{fig:TrapDiag}Beam directions used during $\bhat{y}$ micromotion compensation. For $\bhat{x}$ micromotion compensation the repumper is co-linear with the Doppler cooling beam.}
\end{figure}

Ions with a metastable low-lying D state require a repumping laser to clear out population that decays into it during Doppler cooling.  Here we demonstrate that precise rf correlation in all radial directions can be carried out with this repumping laser placed in an out of plane direction (see Fig.~\ref{fig:TrapDiag}).  This will cause the laser to strike the surface of the trap (and be mostly reflected by the metallic electrodes).  Since this repump laser is in the infrared, unlike the ultraviolet Doppler cooling laser, it is unlikely to cause photoemission and charging of the trap substrate.  Due to the low branching ratio (6\%) of the repumper transition we cannot rely on differing repumping rates at different Doppler shifts to give us a correlation signal.  Instead we operate with high repumper intensity (I$>$50I$_s$ where I$_s\equiv4 \pi \text{hc} \Gamma/\lambda^3$) where a coherent Raman process allows the repumper's Doppler shift to modulate directly the P$_{1/2}$ population, and thus the fluorescence. 

\subsection{Modified detection scheme}

To obtain the optimal parameters for this technique we analyzed the Bloch equations with the inclusion of laser frequency modulation following the method in \cite{ObeTh}.  The equation for the density matrix of atomic states $\tilde{\rho}$ is given by
\begin{equation}
  \label{eqn:bloch}
	\frac{d\tilde{\rho}}{dt}=(M_{0}+\Delta M\cos(\Omega t))\tilde{\rho}
\end{equation}
where the matrix $M_{0}$ is the $8\times8$ Liouvillian describing the Bloch equations of our S$_{1/2}$--P$_{1/2}$--D$_{3/2}$ system (including Zeeman structure) and  $\Omega$ is the trap rf drive frequency. $\Delta M \equiv 2 \pi v_{0} \Delta M_{r}/\lambda_r$ where $\lambda_{r}$ is the repump laser's wavelength, $v_{0}$ is the velocity amplitude in the direction of the laser beam and $\Delta M_{r}$ is the linear change in $M_{0}$ with unit repumper detuning.  The steady state solution of this equation can only contain frequency components at multiples of $\Omega$ and so takes the form

\begin{equation}
  \label{eqn:sum}
	\tilde{\rho}=\sum^{\infty}_{n=-\infty}\tilde{\rho}_{n}e^{-i n \Omega t}.
\end{equation}

\begin{figure}
\includegraphics[width=0.9\columnwidth]{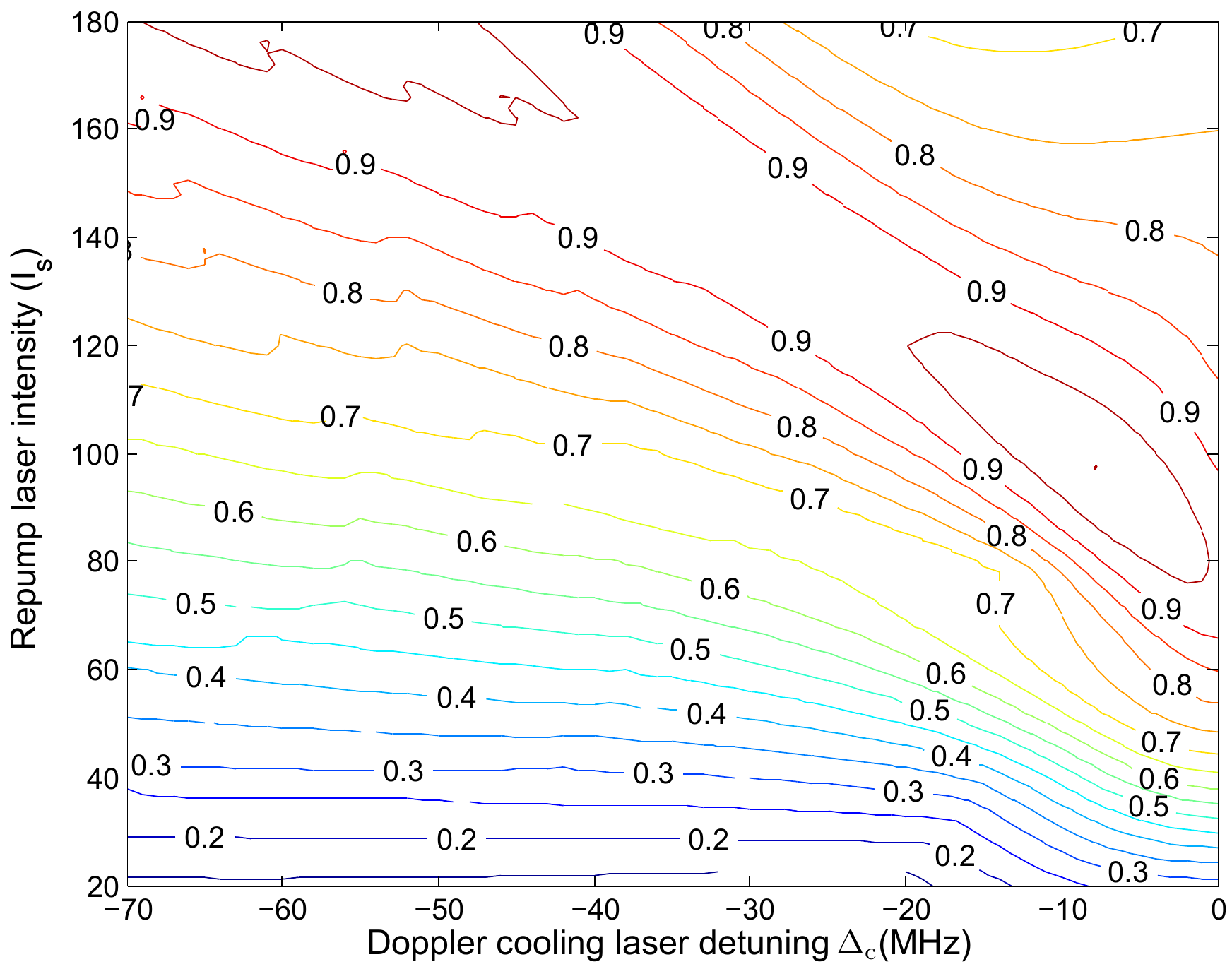}
\caption{\label{fig:sensvsint} Relative sensitivity of micromotion compensation for I$_c$=1.5I$_{s}$ and $\Delta_{r}$ optimized at each point.}
\end{figure}

Substituting Eq.~\ref{eqn:sum} into Eq.~\ref{eqn:bloch} gives the recursion relation
\begin{equation}
	(M_{0}+in\Omega)\tilde{\rho}_{n}+\tfrac{1}{2}\Delta M(\tilde{\rho}_{n+1}+\tilde{\rho}_{n-1})=0
\end{equation}
Terms of $|n|>1$ are negligible in the limit of $\frac{k_{r} v_{0}}{\Omega}\ll 1$ and we obtain solutions determined by
\begin{eqnarray}
	( M_{0}-\tfrac{1}{2}\Delta M(M_{0}^{2}+\Omega^{2})^{-1}M_{0}\Delta M) \tilde{\rho}_0=0\\
	\tilde{\rho}_{\pm1}=-\tfrac{1}{2}(M_0\pm i\Omega)^{-1}\Delta M\tilde{\rho}_0
\end{eqnarray}
which can be evaluated numerically and substituted back into Eq.~\ref{eqn:sum} to give the mean fluorescence and its modulation due to micromotion.  We assume laser linewidths of 500 kHz, $\sigma^{\pm}$ polarizations, a B-field of 1.7G and ignored D$_{3/2}\rightarrow$S$_{1/2}$ decay.  A systematic search of the parameter space of laser intensities and detunings for both the Doppler cooling and repump lasers was carried out to give the optimal parameters.  Figure \ref{fig:sensvsint} shows the maximum micromotion sensitivity as a function of Doppler cooling laser detuning $\Delta_{c}$ and repump laser intensity I$_r$.  Increasing the Doppler cooling intensity I$_c$ causes the peak sensitivity to drop off $\propto(1-0.07\text{I}_c/\text{I}_s)$ but we fix it at I$_c$=1.5I$_{s}$ as this was found experimentally to be a good trade off between sensitivity and the increased signal and cooling given by a higher I$_c$.  The repumper detuning $\Delta_{r}$ is optimized at each point and is found to be in the range of 10--20 MHz to the red of the Doppler cooling laser.  This is fortuitous as it avoids the ion heating that can occur when $\Delta_{r}>\Delta_{c}$ even for $\Delta_{c}<0$ (see Fig.~\ref{fig:microanalysis}).

\subsection{Experimental demonstration}

In the limit of large micromotion the ion sees most of the laser power in motional sidebands which gives a correlation signal that is hard to interpret as the model above is no longer applicable.  Therefore we first coarsely compensate the out-of-plane micromotion by looking at profiles of fluorescence versus $\Delta_\text{r}$, a technique described in \cite{Lis05}.  We find that with $E_\text{dc}\lesssim100~\text{Vm}^{-1}$ the scan gives a well resolved single dark resonance and we can proceed with the correlation method.   Without the tilt potential the rf rail symmetry would mean an $\bhat{x}$ ($\bhat{y}$) compensation field would null $\bhat{x}$ ($\bhat{y}$) micromotion. The tilt voltages introduce some cross-coupling (of order $10\%$) so iterations between $\bhat{x}$ (performed in the standard manner) and $\bhat{y}$ compensation are needed.

\begin{figure}
\includegraphics[width=1\columnwidth]{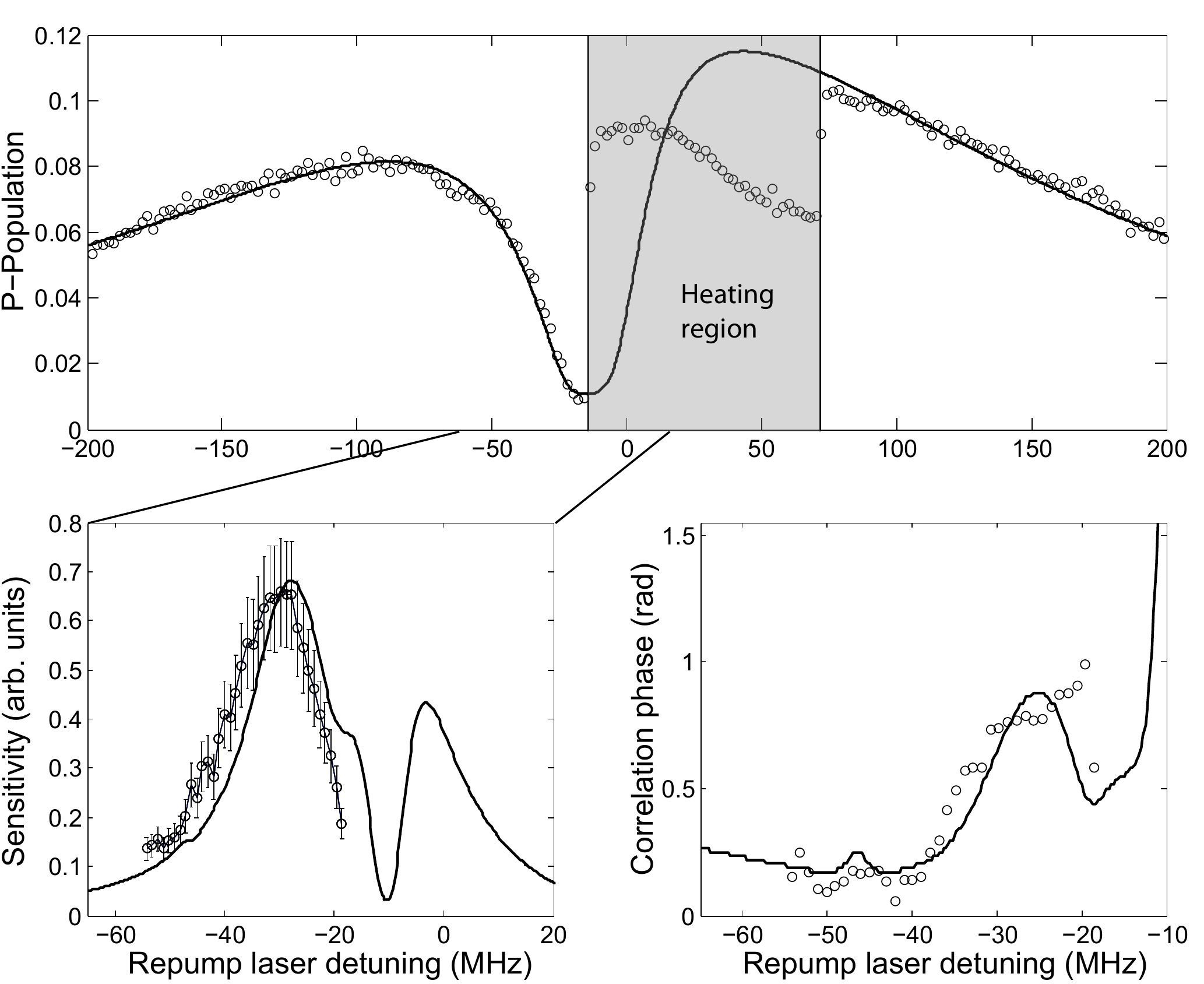}
\caption{\label{fig:microanalysis} Experimental repumper scan (top) and model fitted to give I$_{c}=1.7$I$_s$, I$_{r}=95$I$_s$ and $\Delta_{c}=-14$~MHz.  Predicted micromotion sensitivity based on laser parameters and the experimentally observed sensitivity (bottom left) with vertical scale fitted to the data.  Predicted correlation phase and the experimentally observed phase with the unknown offset fit to the model (bottom right).}
\end{figure}

\begin{figure}
\includegraphics[width=0.9\columnwidth]{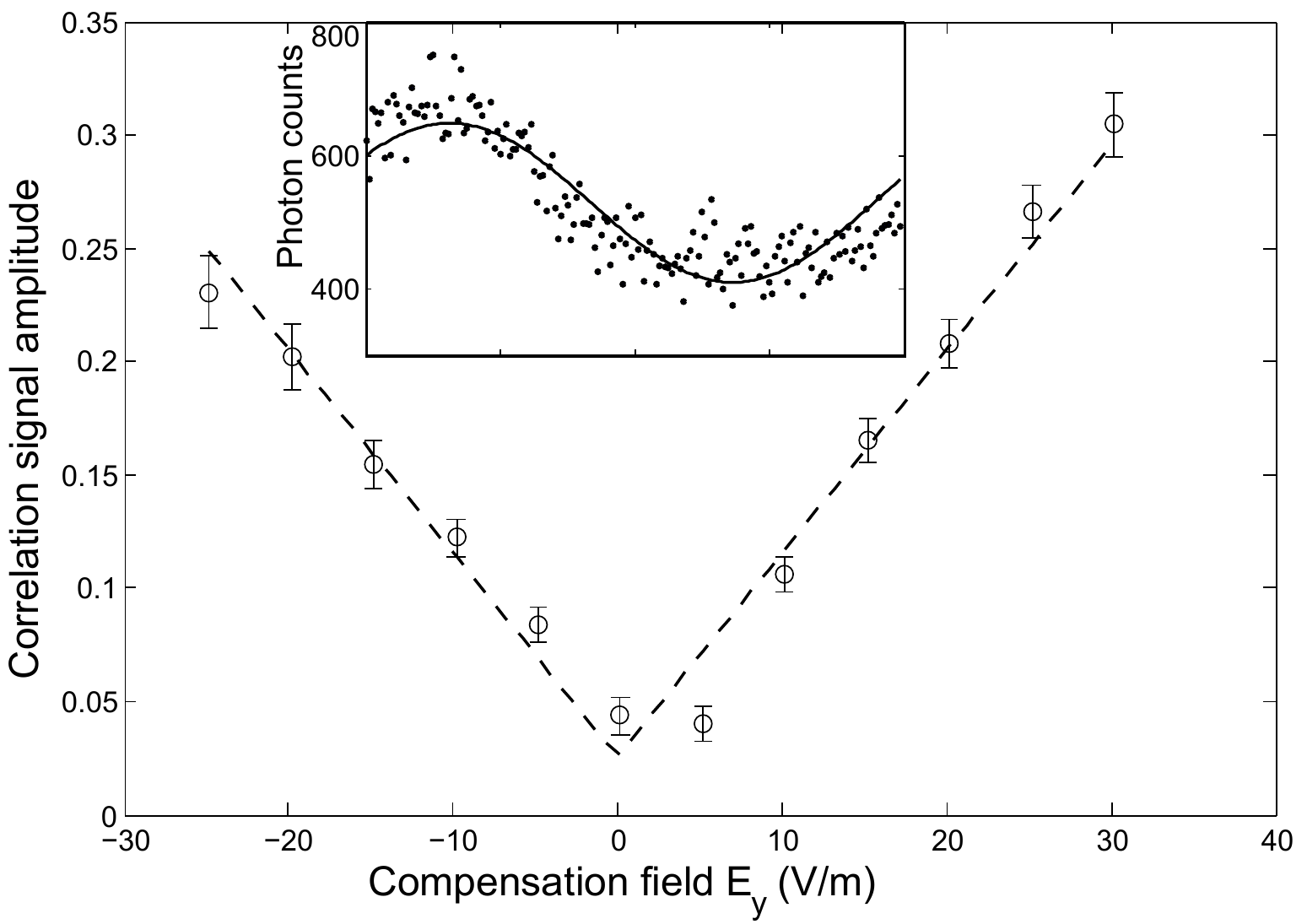}
\caption{\label{fig:comp} Fitted relative amplitude of rf correlation for different $\bhat{y}$ compensation settings and a sample set of correlation data (inset). Laser parameters are as fitted in Figure \ref{fig:microanalysis} with $\Delta_{r}=-28.7$ MHz}
\end{figure}

Using this technique the maximum $\bhat{y}$-direction sensitivity was a relative fluorescence modulation of 0.9(1)\%~per~Vm$^{-1}$ (see Fig.~\ref{fig:comp}) with the laser parameters as in Figure \ref{fig:microanalysis}.  In comparison we obtain $\bhat{x}$-direction sensitivity of 1.3(1)\%~per~Vm$^{-1}$ using the typical co-linear beam arrangement and the Doppler laser detuned to the half-fluorescence point for maximum sensitivity.  Practically this allows us to trim stray fields to below 1~Vm$^{-1}$ in $\bhat{x}$ and 3~Vm$^{-1}$ in $\bhat{y}$.  This corresponds to peak velocity components of 0.1~m/s and  0.3~m/s respectively.  We detect no significant uncompensatable micromotion which would suggest rf phase mismatch \cite{Ber98} or axial micromotion due to an rf-field gradient along $\bhat{z}$.

We observe micromotion compensation fields of up to 1000~Vm$^{-1}$ in both directions with drifts on the order of 10~Vm$^{-1}$ per hour and no noticeable step-change during loading.  In light of this drift rate, a more accurate compensation technique isn't required.  This fairly low drift rate is possibly due to the fact the ion only has direct line of sight to two regions of exposed dielectric, that of the insulating gaps between electrodes and that of the small laser windows.  The former is shielded somewhat by the aspect ratio of the gold electrodes and the latter by the vacuum apparatus and the trap structure.  To check that the out-of-plane repump beam does not cause a change in the micromotion we increased its power to 1~mW ($\approx300\text{I}_s$) and translated it approximately 100~$\mu$m off the ion in the direction of the in-plane beams.  We then monitored the $\bhat{x}$ compensation using those beams whilst alternately turning the out-of-plane beam on and off for several minutes at a time.  There was no detected correlation.  For Doppler cooling, I$_s$ corresponds to $\sim5~\mu$W of laser power (and $20~\mu$W of 389~nm, $80~\mu$W of 423~nm light for photoionization loading). 

\section{\label{sec:heat}Heating Rates}

\subsection{Modified Doppler cooling scheme}

\begin{figure}
\includegraphics[width=0.9\columnwidth]{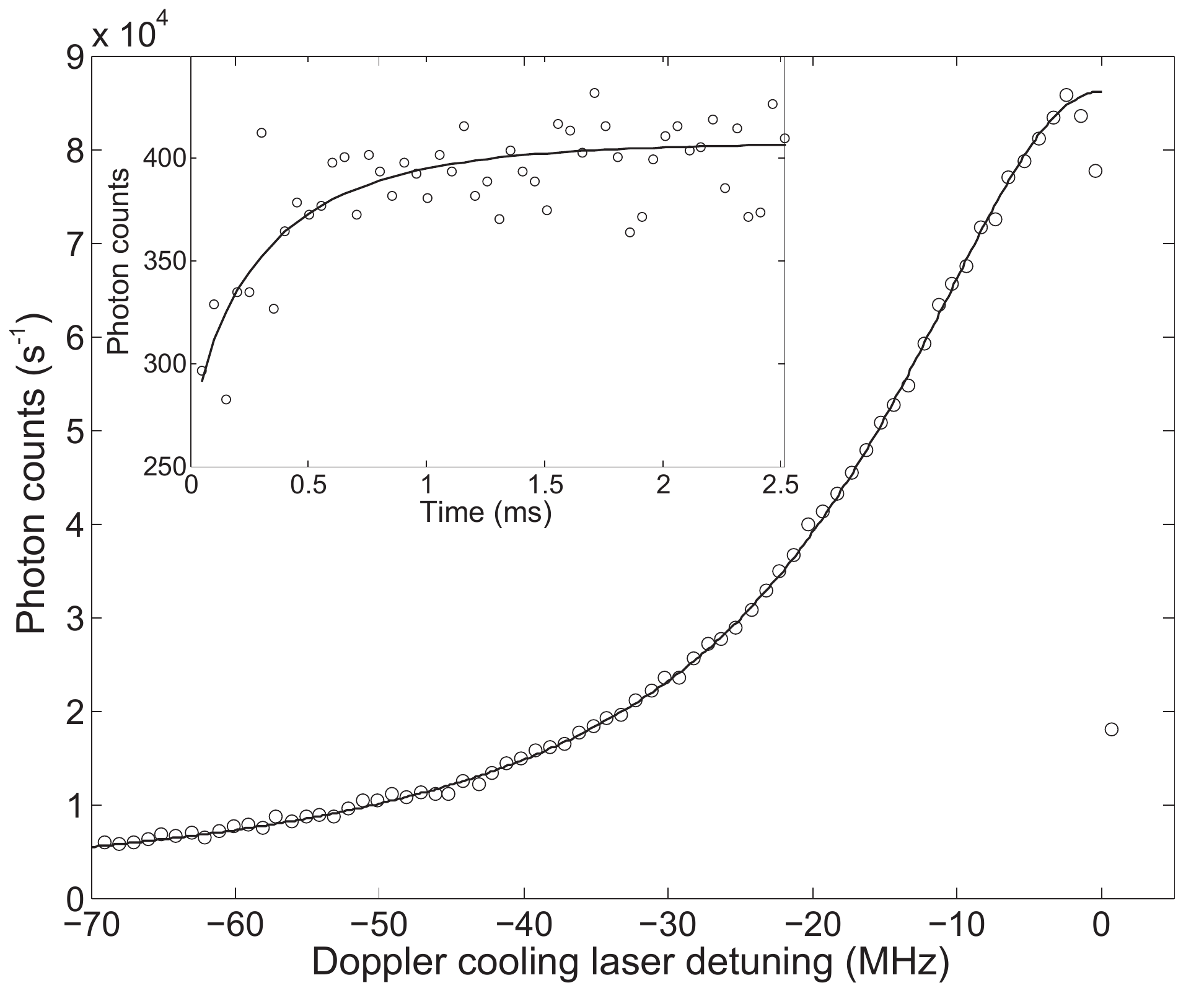}
\caption{\label{fig:recoolscanfit} Fitted fluorescence profile of Doppler cooling laser (s=1.04, $\Gamma$=25.5 MHz) and (inset) a typical Doppler recool experiment (s=1.03, $\Delta$=17.0 MHz).}
\end{figure}

For projective measurement of the S$_{1/2}$--D$_{5/2}$ optical qubit \cite{Mye08} or micromotion compensation we repump decays into the D$_{3/2}$ level using the 866nm transition (see Fig.~\ref{fig:calevs}a).  Coherent dark resonance effects and stimulated emission into the D$_{3/2}$ state complicate the analysis of some experiments such as heating rate measurements and limit the fluorescing P$_{1/2}$ population to 1/4. However, if the P$_{3/2}$--D$_{5/2, 3/2}$ transitions at 850~nm and 854~nm are used to pump the 6\% decay to D$_{3/2}$ back to the S$_{1/2}$ state (see Fig.~\ref{fig:calevs}b) these effects are avoided (There is still a dark resonance if the 854~nm and 850~nm detunings match but for co-propagating beams the difference in the Doppler shifts is small enough that if initially set to different detunings they will not be shifted into a dark resonance by ion motion).   We apply high intensity $(I\sim10^3I_s)$ 850~nm and 854~nm light to power broaden these transitions such that the repump rate is broadly insensitive to Doppler shift; the S$_{1/2}$--P$_{1/2}$ Doppler cooling transition acquires a largely Lorentzian lineshape with deviations due to a small population in the D$_{3/2}$--P$_{3/2}$--D$_{5/2}$ manifold.   Figure~\ref{fig:recoolscanfit} shows fluorescence data fitted by a two-level model where the excited state population is given by
\begin{equation}
	\rho_\text{ee}=\frac{s/2}{1+s+(2\Delta/\Gamma)^2}
\end{equation}
where $s$ is the saturation parameter $s\equiv2|\Omega_\text{Rabi}|^2/\Gamma^2$, $\Delta$ is the Doppler cooling laser detuning and $\Gamma$ is the effective linewidth.  The close agreement of the data to a Lorentzian lineshape means we can interpret the Doppler recool method heating rate measurements exactly as described in \cite{Wes07}.

\subsection{Results}

\begin{figure}
\includegraphics[width=0.9\columnwidth]{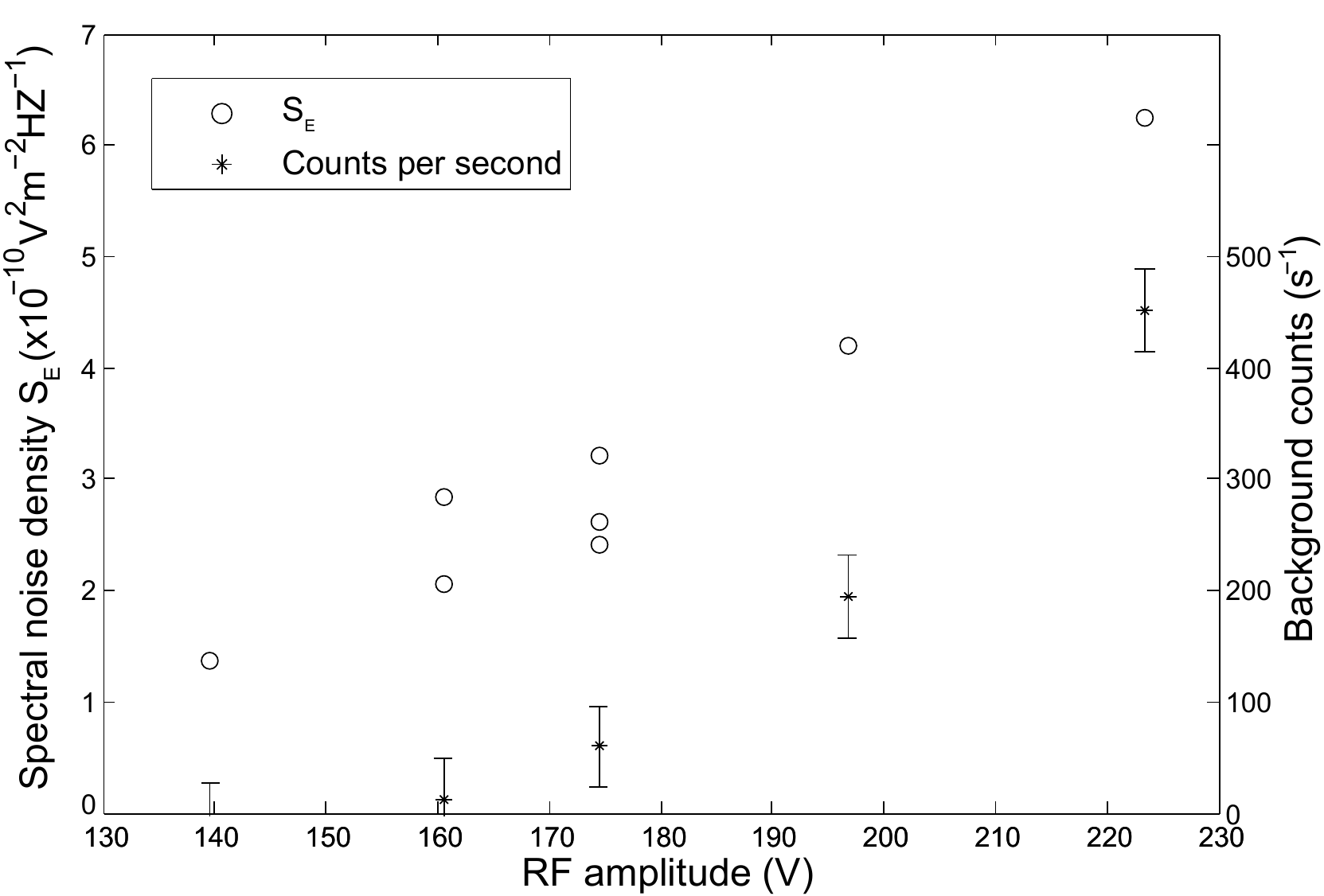}
\caption{\label{fig:heatgraph} Heating rate (spectral noise density of electric field, $S_E$) and background camera counts due to electron emission `glow' from electrodes, both as a function of rf trap voltage.}
\end{figure}

After an ion heating period of one second we measure a temperature of $\bar{n}\approx 5\times10^{4}$ which corresponds to an electric field spectral noise density of \mbox{$S_{E}\approx 2\times10^{-10}~$V$^{2}$m$^{-2}$Hz$^{-1}$} which is comparable to traps of similar size in the literature \cite{Sei06}.  Before this data was fully acquired, a fault in the rf supply caused the trap to arc.  This arcing visibly damaged the surface quality and caused electron emission points to appear at our trap operating voltages.  This emission can be quantified by measuring the difference in background counts with and without the rf supply on, using a CCD camera.  It shows a rapid increase above $\sim150$~V and corresponds with a clear increase in the heating rate (see Fig.~\ref{fig:heatgraph}).  The effect of the decreased surface quality is unknown. The heating rate shows significant day to day shift.

\section{Future Work}

A second, half-sized, version of this trap is currently being fabricated at Sandia National Laboratories using a monolithic silicon process with a view towards on-chip optical system integration.

\begin{acknowledgments}
We thank D. Stick and M. Blain for useful discussions and J. Wesenberg for providing examples of his code.  DTCA would like to thank P. Pattinson and  the photofabrication unit staff. This work was supported by EPSRC (QIP IRC), DTO (contractW911NF-05-1-0297), the European Commission ("SCALA", "MicroTrap") and the Royal Society.
\end{acknowledgments}

\bibliography{trappaper}

\end{document}